\documentclass[preprint2]{aastex}

\newcommand{\be}{\begin{equation}}
\newcommand{\ee}{\end{equation}}
\newcommand{\ba}{\begin{eqnarray}}
\newcommand{\ea}{\end{eqnarray}}
\newcommand{\om}{\Omega}

\newcommand{\vecvsw}{\mathbf{v}_{sw}}
\newcommand{\vecb}{\mathbf{B}}
\newcommand{\vece}{\mathbf{E}}

\slugcomment{ApJ, to be submitted}

\shorttitle{drift-induced deceleration}
\shortauthors{Dalla et al.}

\begin{document}


\title{Drift-induced deceleration of Solar Energetic Particles}


\author{S. Dalla\altaffilmark{1}, M.S. Marsh\altaffilmark{1,2} and T. Laitinen\altaffilmark{1}}
\affil{Jeremiah Horrocks Institute, University of Central Lancashire,
    Preston, PR1 2HE, UK}
\affil{Met Office, Exeter, EX1 3PB, UK}







\begin{abstract}
We investigate the deceleration of Solar Energetic Particles (SEPs)
during their propagation from the Sun through interplanetary space,
in the presence of weak to strong scattering in a Parker spiral configuration,
using relativistic full orbit test particle simulations.
The calculations retain all three spatial variables describing particles'
trajectories, allowing to model any transport across the magnetic field.
Large energy change is shown to occur for protons, due to the combined effect of
standard adiabatic deceleration and a significant contribution 
from particle drift in the direction opposite to that of the solar wind electric field.
The latter drift-induced deceleration is found to have a stronger effect for SEP energies than for galactic
cosmic rays. The kinetic energy of protons injected at 1 MeV is found to be reduced
by between 35 and 90\% after four days, and for protons injected at 100 MeV 
by between 20 and 55\%.
The overall degree of deceleration is a weak function of the
scattering mean free path, showing that, although adiabatic deceleration
plays a role, a large contribution is due to particle drift. 
Current SEP transport models are found to account for drift-induced deceleration
in an approximate way and their accuracy will need to be assessed in future work.
\end{abstract}


\keywords{Solar Energetic Particles, drift, deceleration}

\section{Introduction}

Solar Energetic Particles (SEPs) are commonly measured near 1 AU in the interplanetary medium, 
in most cases at large distances from the region where they were accelerated, in
solar flares and Coronal Mass Ejection (CME) driven shocks. 
Understanding their propagation through space is important to gain information
about the acceleration processes and to test theories of 
charged particle transport in turbulent magnetic fields, relevant to
a variety of astrophysical environments.
Since SEPs are a radiation hazard for satellite infrastructure and 
humans in space, accurate modelling of their propagation is important also to develop
a Space Weather forecasting capability.

During their motion through interplanetary space, SEPs
undergo changes in their kinetic energy. 
At the present time, a well established framework for the description of kinetic energy change of
SEPs exists, based on the theory of adiabatic deceleration within a transport equation, initially developed to
describe galactic cosmic rays. 
In its classic description, adiabatic deceleration results from the cooling of an
isotropic energetic
particle gas as it moves outwards in the expanding solar wind \citep{Par1965}.
\cite{Ski1971} derived a focussed transport equation including adiabatic deceleration for the case when
the plasma convection velocity is at an angle with respect to the magnetic field, retaining pitch-angle as a variable.
\cite{Ruf1995} obtained an alternative formulation for a Parker spiral
geometry, again considering
particles of arbitrary pitch-angle.
Both approaches use a deceleration term within a transport equation.   
The terms derived by \cite{Ski1971} and \cite{Ruf1995} are equivalent for a 
Parker spiral magnetic field.

The majority of modern models of 
SEP propagation include adiabatic deceleration by using either the 
\cite{Ruf1995} or the \cite{Ski1971} formalism
within the focussed transport equation  [e.g.~\cite{Lar1998,Agu2008,Zha2009,Dro2010,Mas2012,Wan2012}]. 
Numerical solution of this equation shows 
that adiabatic deceleration needs to be taken into account when 
studying SEP propagation, as it 
affects the decay phase of events
significantly for proton energies up to about 20 MeV \citep{Ruf1995}. 
Intensity profiles are affected in a complex way that varies in the
different stages of the event \citep{Koc1998}.
\cite{Mas2012} modelled the transport of heavy ions and showed that the effects of adiabatic cooling
are very strong below 1 MeV nucleon$^{-1}$. 
They pointed out that, as a result of deceleration, 
particles measured within a given energy channel of a detector located at 1 AU may have
had a much higher energy when originally injected at the Sun. 

The standard focussed transport approaches \citep{Ski1971, Ruf1995} assume that energetic particles remain tied to magnetic
field lines during their propagation, so that only a single spatial variable, the
distance travelled along a magnetic field line, needs to be retained to describe the particle's trajectory.
This is a common assumption in many studies involving SEP modelling and data analysis.   

However in two recent publications we pointed out that drifts due to the gradient and curvature
of the Parker spiral magnetic field are important for SEPs \citep{Mar2013, Dal2013}.
This result is based on full orbit test particle simulations of SEPs in a Parker spiral configuration (\cite{Mar2013}, Paper I) as well as on analytical expressions for the particle drifts (\cite{Dal2013}, Paper II). 
As a result of drift, particles experience transport in longitude and latitude, perpendicular to the magnetic field line on which they were originally injected.
The scattering conditions, described by a mean free path $\lambda$, do not strongly affect the amount of drift experienced. In fact, significant
curvature drift is found to be present even when particles propagate scatter-free.

In this paper we examine an important consequence of drifts for SEPs: the strong
particle deceleration due to drift motion in the direction opposite to that of the solar wind electric field, studied here for the case of protons. The electric field is present in the fixed (observer) frame due to the outward propagation of the solar wind plasma. Drift-induced deceleration has been studied  previously 
in the context of galactic cosmic rays \citep{Kot1979, Bar1975}, but has not been discussed for SEPs up
to the present time.
We show that, unlike adiabatic deceleration, drift-induced deceleration is present also in the absence of scattering. We evaluate whether the standard focussed transport equation formalisms  \citep{Ski1971, Ruf1995}
describe the effects of drift-induced deceleration on the particle distribution function.

\section{Energy change for SEPs as derived from test particle simulations}\label{sec.test_simul}

In Paper I, the propagation of SEPs in a Parker spiral configuration in the presence of 
scattering was studied by means of a relativistic full orbit test particle code.
The output of the simulations showed that, rather than remaining confined within the magnetic
field lines delimiting the injection region at the Sun, particles experience significant 
drift (see e.g. Figures 3 and 5 of Paper I).
The amount of drift, measured in a local Parker spiral coordinate system 
as the displacement 
from the magnetic field line through the initial position, 
increases with energy, i.e.~it is much larger for 100 MeV  
than for 1 MeV  protons. It is particularly large for partially ionised
heavy ions due to proportionality of grad B and curvature drift velocities to the charge to mass
ratio.

Here we examine the results of simulations similar to those presented in Paper I,
focussing on the change in kinetic energy experienced by the particles during their
propagation.
We follow four initially monoenergetic populations, each consisting of 1000 protons, characterised by values of the initial kinetic energy $K_0$ of 1 MeV, 10 MeV, 100 MeV and 1 GeV.
The particles are injected from an 8$^{\circ}$$\times$8$^{\circ}$ region at 1 solar radius and 
centered at latitude $\delta$=20$^{\circ}$. Their initial velocity directions are randomly chosen in a semi-hemisphere in velocity space pointing away from the Sun. 
Particles are scattered with an average frequency defined by a scattering mean free path $\lambda$. During each scattering event the particle's velocity is
randomly reassigned from an isotropic distribution. Particles are scattered in the solar wind frame
and in this frame energy is assumed to be conserved during a scattering event.  This is a reasonable approximation since the majority of scattering events in our simulations take place far from the Sun. For comparison, scatter-free runs are also performed. Other parameters of the simulations have been described in Paper I.

The Parker spiral magnetic field used in the simulations is unipolar (pointing outwards from the Sun, as was assumed in the simulations of Paper I), i.e.~the
presence of the heliospheric current sheet is not modelled.
The Lorentz equation of motion for each particle is solved  
in the fixed frame, in which the Sun is rotating 
with angular velocity $\Omega$ and the solar wind is moving radially outward at constant speed $v_{sw}$. This
is the frame which is relevant for comparison of simulation results with observations
taken by spacecraft.
 
\begin{figure}
\epsscale{.80}
\plotone{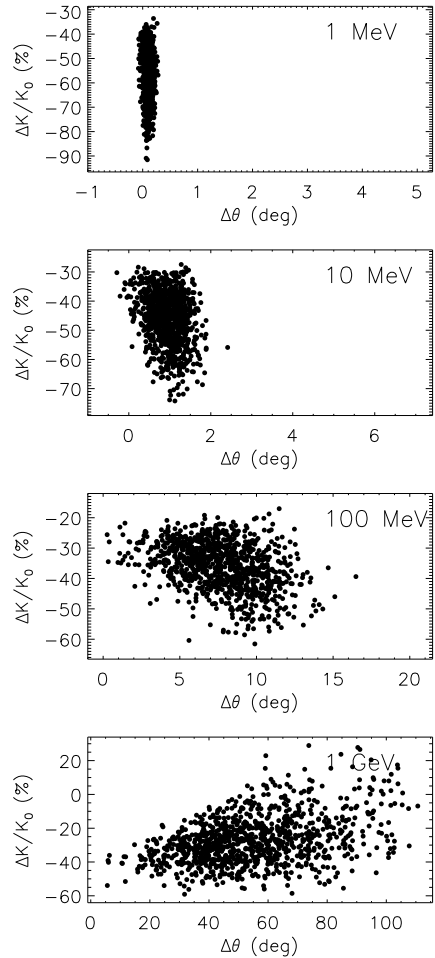}
\caption{Percentage change in kinetic energy as a function of colatitude change $\Delta \theta$, four days after injection, for protons of four initially monoenergetic populations with initial energy $K_0$ indicated in each panel. Particles are injected from an 8$^{\circ}$$\times$8$^{\circ}$ region at latitude $\delta$=20$^{\circ}$ and the mean free path is $\lambda$=1 AU.\label{fig.kinenergy_perc_change}}
\end{figure}

Figure \ref{fig.kinenergy_perc_change} shows the kinetic energy change experienced by particles of the four initially monoenergetic populations (one for each panel, with initial energy $K_0$ as labelled) at time $t_f$=102 hours, plotted as a function of the change  in colatitude experienced by each particle. 
The mean free path is $\lambda$=1 AU at all energies.
Kinetic energy change $\Delta K$=$K$$-$$K_0$ is shown as percentage of the initial energy $K_0$, as a function of $\Delta \theta$=$\theta$$-$$\theta_0$ where $\theta$ is the particle's colatitude and $\theta_0$ its initial value. Particles in all four populations are seen to have experienced strong deceleration over the time of the simulation, apart from a minority of those with $K_0$=1 GeV, whose kinetic energy increases.

It should be noted that the scale on the $x$-axes of the panels of Figure \ref{fig.kinenergy_perc_change} is different for the four populations, showing that the 1 MeV particles have 
a much smaller $\Delta \theta$ compared to 1 GeV particles. 
The observed change in colatitude
is caused by drift due to the curvature and gradient of the 
Parker spiral magnetic field (Paper I, Paper II), and the amount of drift increases with particle
energy.

\begin{figure}
\epsscale{.8}
\plotone{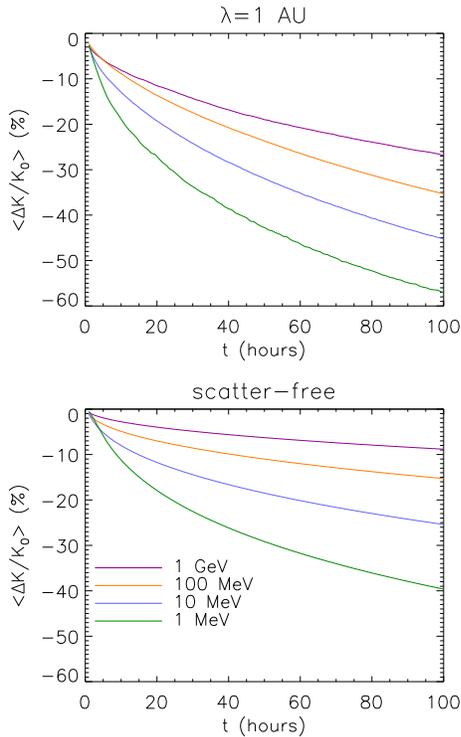}
\caption{Time variation of $\Delta K / K_0$  averaged over each initially monoenergetic population with $K_0$= 1 MeV, 10 MeV, 100 MeV and 1 GeV. The top panel is for $\lambda$=1 AU and the bottom panel for the
scatter-free case. Other parameters of the simulations are as in Figure \ref{fig.kinenergy_perc_change}.
\label{fig.kinenergy_time_change}}
\end{figure}

Figure \ref{fig.kinenergy_perc_change} shows a trend for the deceleration to increase with $\Delta \theta$ for the cases where  $K_0$= 1 MeV, 10 MeV and 100 MeV, but not for the case $K_0$= 1 GeV.

In Figure \ref{fig.kinenergy_time_change} the kinetic energy change averaged over each of the four particle populations, is plotted versus time. The top panel shows the case $\lambda$=1 AU, corresponding
to the same simulations shown in Figure \ref{fig.kinenergy_perc_change}, while the bottom panel 
presents the time dependence of energy change in the scatter-free case.
Figure \ref{fig.kinenergy_time_change} shows that the decrease in  kinetic energy is fast during the first 
$\sim$20 hours of the propagation, and continues to take place more slowly at later times. 
Deceleration is seen also in the absence of scattering, although it is of a reduced
magnitude compared to the $\lambda$=1 AU scattering case.

\section{Deceleration processes}\label{sec.kinenergy}

\subsection{Energy change due to drift in the electric field potential} \label{sec.driftinduced}

As energetic particles propagate through the interplanetary medium, in the fixed, non-rotating reference frame they are subject to the electric field associated with the outwardly moving solar wind. The solar wind electric field is given by:
\be
\vece = -  \frac{\vecvsw \times \vecb}{c}   \label{eq.electricfield1}
\ee
where $\vecvsw$ is the solar wind velocity, $\vecb$ is the interplanetary magnetic field (IMF) and $c$ is the speed of light.

We consider $\vecb$ to be a unipolar Parker spiral, and the solar wind flow to be radial, uniform over the Sun's source surface and time independent. With these assumptions, the expression for the electric field takes the form \citep{Bur1968, Win1968}:
\be
\vece = - \frac{A}{c} \, \om  \, B_0 \, \frac{r_0^2}{r} \, \sin{\theta} \,\, \mathbf{e}_{\theta}  \label{eq.etheta}  \\
\ee
where $\mathbf{e}_{\theta}$ is the standard unit vector in a ($r,\theta,\phi$) spherical coordinate system centered on the Sun,
with $r$ the radial distance, $\theta$ the colatitude and $\phi$ the longitude. Here $B_0$ is the magnitude of the magnetic field at a reference radial distance $r_0$ and
$\om$=2.87$\times$10$^{-6}$ rad s$^{-1}$ is the solar rotation rate.
$A$=1 when $\vecb$ points outward from the Sun and and $A$=$-$1 when $\vecb$ points inwards. 

The potential associated with the electric field of Eq.~(\ref{eq.etheta}) is given by:
\be
\Phi = - \frac{A}{c} \, \om \, B_0 \, r_0^2  \, \cos{\theta} \label{eq.epotential}
\ee

In the absence of other effects causing deceleration, the kinetic energy change due to motion in the direction of the solar wind electric field is given by:
\be
\Delta K_{(\Phi)} = \frac{q \, A}{c} \, \om \, B_0 \, r_0^2 \, (\cos{\theta} - \cos{\theta_0})  \label{eq.delta_ek_potential}
\ee
with $q$ the particle charge, $\theta_0$ its initial colatitude and the subscript ${(\Phi)}$ indicating that this is the energy change associated with the potential of Eq.~(\ref{eq.epotential}).
Expressions equivalent to Eq.~(\ref{eq.delta_ek_potential}) were derived by \cite{Kot1979}.

If particles do not move in colatitude, $\theta$=$\theta_0$ and
the kinetic energy change given by Eq.~(\ref{eq.delta_ek_potential}) is zero.
However Papers I and II demonstrated that curvature and grad-$B$ drift combine to produce a significant component of drift in ${\theta}$, which, from Eq.~(\ref{eq.delta_ek_potential}), will cause kinetic energy change.
The drift in colatitude is present also in the absence of scattering.

Eq.~(\ref{eq.delta_ek_potential}) can be rewritten as:
\be
\Delta K_{(\Phi)} = A\,  \mbox{sign}(q) \, \xi_0 \, (\cos{\theta}-\cos{\theta_0})  \label{eq.delta_ek_potential2}
\ee
where:
\be  \label{eq.xi0_def}
\xi_0= \frac{|q|}{c} \,  \om \, B_0 \, r_0^2  
\ee
For a proton or an electron, using the same values for $B_0$ and $r_0$ used in Papers I and II, we obtain that $\xi_0$=247 MeV.

Paper II showed that for the case of a positive polarity of the Parker spiral IMF ($A$$>$0) and particle initial positions not far from the heliographic equatorial plane, ions drift downwards (increasing colatitude) and electrons upwards (decreasing colatitude). Hence from Eq.~(\ref{eq.delta_ek_potential2}) both ions and electrons will experience a negative $\Delta K_{(\Phi)}$, i.e.~a deceleration, as a result of drift motion.

For a negative polarity configuration ($A$$<$0) protons near the equatorial plane drift upwards and electrons downwards, so that the effect of the drift is again deceleration for ions and electrons.

The expression for the energy change (Eq.~(\ref{eq.delta_ek_potential})) does not involve the initial particle energy explicitly, however the latter is an important parameter in determining how much the particle drifts in colatitude, i.e.~the value of $\cos{\theta}-\cos{\theta_0}$.

\begin{figure*}
\epsscale{1.90}
\plotone{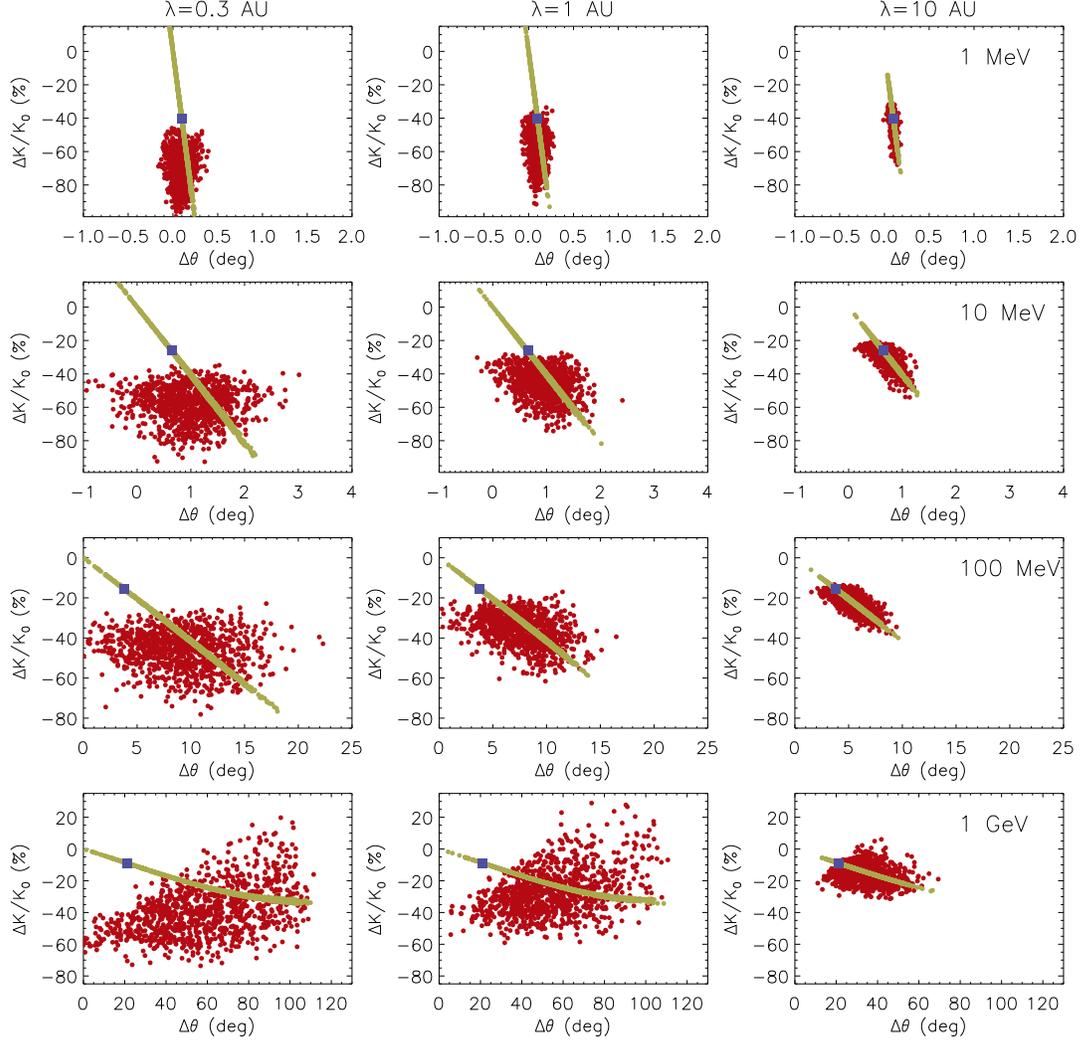}
\caption{Percentage change in kinetic energy as a function of $\Delta \theta$, 
four days after injection, 
for three values of the scattering mean free path $\lambda$ and varying initial kinetic energy.
The red dots are from simulations including both adiabatic deceleration and drift-induced deceleration,
the green dots are for simulations without adiabatic deceleration, i.e. scattering taking place in the fixed frame. The blue square represents the scatter-free case.
\label{fig.lambda_dependence}}
\end{figure*}

\subsection{Energy change due adiabatic deceleration}

In standard transport theory, heliospheric energetic particles experience cooling due to the expansion of the solar wind, which carries the turbulence that produces particle scattering. Formalism to describe this adiabatic deceleration within a transport equation was derived by \citet{Par1965} for the case of an isotropic particle distribution function and radially directed magnetic field.

Particle scattering in the solar wind frame is essential for adiabatic deceleration. In the absence of scattering, this process does not take place.

\citet{Ruf1995} generalised the classic treatment to the case of particles of arbitrary pitch angle and a Parker spiral configuration, pointing out how the combination of focussing and scattering produces a deceleration in the solar wind frame.
The rate of deceleration for a single particle is given by \citep{Ruf1995}:
\be \label{eqn.ad_decel}
\frac{d p'}{dt}= - p' v_{sw}  \left[  \frac{\sec{\psi}}{2 L(z)} (1-\mu'^2) + \cos{\psi}\frac{d }{dr}\sec{\psi} \mu'^2 \right]
\ee
where $p'$ and $\mu'$ are the particle's momentum and pitch angle in the solar wind frame, $\psi$ is the angle between the Parker spiral magnetic field line and the radial direction and $L(z)$ is the focussing length. Here $z$ is the only spatial variable retained within the model, describing the distance travelled along a magnetic field line.

From Eq.~(\ref{eqn.ad_decel}), the characteristic time of deceleration $\tau$=$-$$(1/p'\, dp'/dt)^{-1}$ depends only on the geometrical parameters, including the colatitude, via the angle $\psi$ and $L$, and on the particle pitch angle, i.e. it does not depend explicitly on the particle's kinetic energy nor species. 
However the latter parameters influence the overall trajectory and amount of time that a particle spends at a given location, and as a result the overall energy change experienced.
If the scattering mean free path is small, particles tend to spend a longer time in regions near the Sun, where adiabatic deceleration
is stronger, so that a lower value of $\lambda$ gives rise to more deceleration associated with this process  \citep{Par1965}.

An adiabatic deceleration formalism alternative to Eq.~(\ref{eqn.ad_decel})
had been derived earlier by \cite{Ski1971}, for a generic magnetic field.
For the case of a Parker spiral field, it can be shown analytically that the  \cite{Ski1971} expression  
takes the same form as the \citet{Ruf1995} one, so that the two formulations
are equivalent in this case, as noted in the latter paper.

\section{Interpretation of simulation results}

Having introduced the concepts of drift-induced and adiabatic deceleration we can now interpret the results shown in 
Figures  \ref{fig.kinenergy_perc_change} and \ref{fig.kinenergy_time_change}. 
Our simulations do include adiabatic deceleration because scattering is taking place in the solar wind frame. They also include 
drifts which produce changes in colatitude and thus energy change. 

In Figure  \ref{fig.kinenergy_perc_change}, the observed trend for deceleration to increase with $\Delta \theta$ (as seen for  $K_0$= 1 MeV, 10 MeV and 100 MeV) is consistent with drift-induced deceleration playing a significant role.
The fact that deceleration is present also in the scatter-free case (Figure \ref{fig.kinenergy_time_change}, bottom panel), when no adiabatic deceleration is taking place, confirms this statement.

Low energy SEPs experience very little drift, however their initial kinetic energy is much smaller than $\xi_0$ (given by Eq.~(\ref{eq.xi0_def})), 
the magnitude of the largest possible deceleration associated with drift processes.
Considering Eq.(\ref{eq.delta_ek_potential2}), one can see that
even a small $\Delta \theta$ can produce a significant percentage change in their energy, the behaviour shown in Figure \ref{fig.kinenergy_perc_change} (e.g.~for $K_0$=1 MeV). 
SEPs of initial energy 1 GeV typically drift by tens of degrees, however their drift-induced deceleration as percentage of initial energy is much less than for lower energy particles due to the larger ratio $K_0/\xi_0$. As one considers particles of even higher energy, drift-induced deceleration becomes less significant.

The deceleration seen in Figures  \ref{fig.kinenergy_perc_change} and \ref{fig.kinenergy_time_change} is a combination of adiabatic and drift-induced deceleration.
However it is possible to \lq switch off\rq\ adiabatic deceleration by scattering particles in the fixed frame, rather than in the solar
wind frame. When this is done, particle energy is conserved in the fixed frame during a scattering event,
while a change in pitch angle does result, influencing the particle's trajectory.
Scatter-free simulations represent an extreme case: 
here no adiabatic deceleration takes place due to the absence of 
scattering events and particles propagate together as a beam. 
Drift and the deceleration it produces cannot be eliminated in our simulations.

Figure \ref{fig.lambda_dependence} shows the output of simulations carried out in three very different scattering conditions, each corresponding to a column in the figure.
Red points are the results for the cases with adiabatic deceleration (scattering carried
out in the solar wind frame). Points in the central column ($\lambda$=1 AU) 
are the same as in Figure \ref{fig.kinenergy_perc_change}, and the
columns on the left and right show the results for $\lambda$=0.3 AU and 10 AU respectively.
The light green points
show the energy change observed when scattering is carried out in the fixed frame, i.e.~no adiabatic deceleration is occuring,
for the same values of the mean free path. 
The blue squares represents the deceleration observed in the scatter-free case: here all particles experience
the same $\Delta \theta$ due to drift, resulting in a single point in the plots and no adiabatic deceleration.

Considering first of all the red points in Figure \ref{fig.lambda_dependence}, one can see that as the level of scattering increases
(i.e.~moving from right to left in the figure),
more deceleration is observed and the range of percentage change in $\Delta K$ increases. 
The amount of drift, represented by the range of $\Delta \theta$ values also increases, as was noted in Paper I.
Considering that in our simulations the mean free path values
vary by almost 2 orders of magnitude, from 10 AU to 0.3 AU, the amount of deceleration appears to be only a weak function of the
mean free path. 

When no adiabatic deceleration is present (light green points), the amount of deceleration observed has much less scatter and follows a line which is found to coincide exactly with that 
predicted by Eq.~(\ref{eq.delta_ek_potential2}), i.e.~the observed energy change is purely drift-induced deceleration.
It is also apparent that any effects due to large finite Larmor radius are small in our simulations, as the data points for this case lie
along a single line in   Figure \ref{fig.lambda_dependence}, without significant scatter.

We can now compare the results for the cases with adiabatic deceleration, without it and the scatter-free situation.
Looking at the 
low scattering regime ($\lambda$=10 AU), it is apparent that the addition of adiabatic deceleration broadens the distribution around the line predicted by Eq.~(\ref{eq.delta_ek_potential2}), and the
effect becomes more marked for stronger scattering. Switching on adiabatic deceleration eliminates the low values of $|\Delta K / K_0|$ which 
are present in the case of purely drift-induced deceleration (e.g.~for $K_0$=1 MeV values of $|\Delta K / K_0|$$<$30$^{\circ}$). 
In the strong scattering case, adiabatic deceleration is more efficient, resulting in a distribution
of $\Delta K  / K_0$ values which does not show a clear dependence on $\Delta \theta$, unlike for the $\lambda$=10 AU case.
The weak dependence of the overall deceleration on the value of the mean free path demonstrates that drift-induced deceleration plays a very significant role. 

Regarding the reason why a small minority of particles with $K_0$=1 GeV show acceleration (Figure \ref{fig.kinenergy_perc_change}, bottom panel), we note that it is possible for the process responsible for adiabatic deceleration to produce acceleration in the fixed frame in some cases (as can be seen from the qualitative description accompanying Figure 1 of \citet{Ruf1995}). 

In the simulation discussed so far, particles were injected at 1 solar radius. We have also carried out simulations with injection at 40 solar radii for $\lambda$=1 AU and the deceleration observed is very similar to that shown in Figure \ref{fig.kinenergy_perc_change}.
Therefore the effects describe above apply both to particles energised in solar flares and to those accelerated in CME driven-shocks at distances up to at least  40 solar radii.

\section{Discussion and conclusions}

Our relativistic full orbit test particle simulations of proton propagation in the Parker spiral have allowed us to analyse, for the first time,
the deceleration associated with drift motion opposite to the solar wind electric field.
We showed that the overall deceleration experienced by SEPs is due to contributions from 
adiabatic deceleration (which requires scattering) and drift-induced deceleration (which is present
also in the absence of scattering).

We found that energy changes associated with the combined processes of drift-induced and adiabatic deceleration are very large:
for a scattering mean free path of $\lambda$=1 AU,  $\Delta K / K_0$ at 100 hours after injection ranges between $-$35\% and $-$90\% 
for particles with $K_0$=1 MeV, between  $-$30\% and $-$70\% for $K_0$=10 MeV and between  $-$20\% and $-$55\% for $K_0$=100 MeV. 

It is interesting to note that while the magnitude
of drift, as measured by the change in colatitude,  is very small for protons at $\sim$1 MeV, its effect
on particle energy is large, 
meaning that drift effects cannot be ignored even at these low energies.

With reference to the very large values of deceleration obtained from our simulations for  particles of initial energy of 1 MeV, 
e.g.~values close to 100\%, the test particle approximation breaks down once the particle energy has reached values close enough to
that of solar wind particles, so that a different modelling approach is required.

Deceleration is fast during the first $\sim$20 hours following injection and continues at a slower rate in the subsequent hours.
The fact that it is taking place also for scatter-free propagation means that it cannot be ascribed to standard adiabatic
deceleration only.
In addition, the simulations with scattering taking place in the fixed frame (i.e.~with scattering present but 
adiabatic deceleration \lq switched off\rq) demonstrate the large magnitude of drift-induced deceleration.

The magnitude of the constant $\xi_0$ appearing in the expression for the solar wind electric potential (Eq.~(\ref{eq.delta_ek_potential2})) determines the particle energy 
range over which the effect of drift-induced deceleration is most prominent. 
It appears that for protons this is the range between $\sim$1 and a few hundred MeV, which is the typical SEP energy range.
At proton energies larger than 10 GeV the energy change becomes negligible, therefore its importance for galactic cosmic rays is for modulated cosmic rays in the solar wind with energies in the $\sim$1 MeV -- 1 GeV range. 

We also emphasize that drift additional to that due to the large scale Parker spiral field and  any other process which is able to produce transport in the latitudinal direction will cause
changes in kinetic energy.

Is drift-induced deceleration included within the standard transport equations \citep{Ski1971, Ruf1995} used in SEP modelling, data analysis and Space Weather forecasting? Answering this question requires 
analysis of the derivation of the focussed transport equation \citep{Ski1971}. This equation describes the evolution of the distribution function of the particles' guiding centres and 
is obtained by averaging the Vlasov equation over gyrophase (e.g.~\citet{Zan2014}). 
\citet{LeR2009} showed that the focussed transport equation is equivalent to the guiding centre kinetic equation \citep{Kul1983}. The latter is derived by means of the standard assumptions of first order 
guiding centre theory, including the assumption that the grad B and curvature drift velocities are much smaller than the {\bf E}$\times${\bf B} drift velocity.
As shown by \citet{LeR2009}, the guiding centre kinetic equation includes a contribution from grad B and curvature drift in its energy change term, while the latter drifts are not included in the 
spatial convection term. This means that the energy change associated with drift is taken into account in an approximate way, with the deceleration rate being calculated along the \lq undrifted\rq\ orbit, at constant colatitude. 
To our knowledge, all current SEP transport models (e.g.~\citet{Zha2009}, \citet{Mas2012}) make use of this approximation, which we expect to be fairly accurate for low energy SEPs but less accurate for high energy ones, due to their substantial latitudinal drift. 
A more precise solution would be obtained by solving a transport equation that includes the grad B and curvature drifts within the spatial convection term, i.e. by solving eqs.~(30)--(34) of \citet{Zha2006}.

It should also be noted that in Paper II we showed that the grad B and curvature drift velocities in the Parker spiral can be of the same
order of magnitude as the {\bf E}$\times${\bf B} drift velocity, and in some cases even larger. Therefore the standard guiding centre theory assumption
that the {\bf E}$\times${\bf B} drift is the dominant one may not be valid for the case of heliospheric propagation.

We conclude that the extent to which the  drift-induced deceleration observed in our test particle simulations is accounted for by SEP models based on the focussed transport equation needs to be evaluated, and this will be the subject of future work.

Strong SEP deceleration such as the one we reported, has an impact on the analysis of measurements, because it introduces a scenario
where particles detected in a given energy channel at 1 AU actually left the Sun having a much higher energy, as was emphasized 
by \cite{Mas2012} for heavy ions in the $\sim$MeV/nucleon range. Our results show that this may be taking place for protons
in the 10-100 MeV range too.  

Overall, our results demonstrate that drift-induced deceleration is an important process in SEP propagation and needs accurate inclusion within SEP modelling and data analysis.



\acknowledgments

This work has received funding from the European Union Seventh
Framework Programme (FP7/2007-2013) under grant agreement n.~263252
[COMESEP] and from the UK Science and Technology Facilities Council (STFC) (grants ST/J001341/1 and
ST/M00760X/1). SD acknowledges support from ISSI through funding for the International Team on \lq Superdiffusive transport in space plasmas and its influence
on energetic particle acceleration and propagation\rq. 
We thank R.~Vainio for useful discussions.

\bibliographystyle{apj}
\bibliography{energy_biblio}


\clearpage

\end{document}